\documentstyle[11pt,aaspp4]{article}

\begin{document}

\title{Deciphering Core Collapse Supernovae: Is Convection the Key? 
\\ I. Prompt Convection}

\author{A. Mezzacappa\altaffilmark{1,2}, 
A. C. Calder\altaffilmark{1,3}, 
S. W. Bruenn\altaffilmark{4}, 
J. M. Blondin\altaffilmark{5}, 
M. W. Guidry\altaffilmark{1,2},
M. R. Strayer\altaffilmark{1,2}, and 
A. S. Umar\altaffilmark{3}}

\altaffiltext{1}{Theoretical and Computational Physics Group, Oak Ridge National
Laboratory, Oak Ridge, TN 37831--6354}
\altaffiltext{2}{Department of Physics and Astronomy, University of Tennessee,
Knoxville, TN 37996--1200}
\altaffiltext{3}{Department of Physics and Astronomy, Vanderbilt University,
Nashville, TN 37235}
\altaffiltext{4}{Department of Physics, Florida Atlantic University, Boca Raton, FL 33431--0991}
\altaffiltext{5}{Department of Physics, North Carolina State University, 
Raleigh, NC 27695--8202}


\begin{abstract}
We couple two-dimensional hydrodynamics to detailed one-dimensional 
multigroup flux-limited diffusion neutrino transport to investigate
prompt convection in core collapse supernovae. Our initial conditions, 
time-dependent boundary conditions, and neutrino distributions for 
computing neutrino heating, cooling, and deleptonization rates are 
obtained from one-dimensional simulations that implement multigroup 
flux-limited diffusion neutrino transport and one-dimensional hydrodynamics. 
The development and evolution of prompt convection and its ramifications
for the shock dynamics are investigated for both 15 and 25 M$_{\odot}$ 
models, representative of the two classes of stars with compact and 
extended iron cores, respectively. In the absence of neutrino transport, 
prompt convection develops and dissipates on a time scale $\sim$15 ms for 
both models. Prompt convection seeds convection behind the shock, which 
causes distortions in the shock's sphericity, but on the average, the 
shock radius is not boosted significantly. In the presence of neutrino 
transport, prompt convection velocities are too small relative to 
bulk inflow velocities to result in any significant 
convective transport of entropy and leptons. A simple analytical model 
supports our numerical results, indicating that the inclusion of transport 
reduces the convection growth rates and asymptotic velocities by factors 
of 4--250. 
\end{abstract}

\section {\bf Introduction}

Ascertaining the core collapse supernova explosion mechanism is a long-standing
problem in astrophysics. The current paradigm begins with the collapse of a 
massive star's iron core and the generation of an outwardly propagating
shock wave that results from core rebound. Because of dissociation and neutrino losses 
the shock stagnates. This sets the stage for a shock reheating mechanism 
whereby neutrino energy deposition via electron neutrino and antineutrino 
absorption on nucleons behind the shock reenergizes it (Bethe \& Wilson 1985; 
Wilson 1985); however, no recent numerical simulations of shock reheating and 
the subsequent shock evolution produce explosions unless the neutrino 
luminosities or the postshock neutrino heating efficiencies are boosted 
by some other mechanism. One mechanism may be convection (Wilson \& Mayle 
1993, Herant et al. 1994, Burrows et al. 1995, Janka \& M\"{u}ller 1996). 

There are at least three modes of convection that may develop during
the shock reheating phase. (1) Prompt convection near and below the 
``neutrinospheres,'' which is the subject of this Letter and 
which may occur immediately after the formation of the shock (Burrows \&
Fryxell 1993, Bruenn \& Mezzacappa 1994, Janka \& M\"{u}ller 1996). 
(2) Doubly diffusive instabilities below the neutrinospheres [e.g., 
``neutron fingers'' (Wilson \& Mayle 1993); however, see Bruenn \& 
Dineva 1996]. (3) Neutrino-driven convection below the shock (Herant 
et al. 1992, Miller et al. 1993, Herant et al. 1994, Burrows et al. 1995, 
Janka \& M\"{u}ller 1996, Mezzacappa et al. 1996, Calder et al. 1996b). 

Prompt convection may initially be driven by negative entropy gradients 
imprinted on the core by the weakening shock and may shortly thereafter 
be sustained by the negative lepton gradient that arises near and below 
the electron neutrinosphere as a result of electron neutrino losses. 
Based on multidimensional hydrodynamics and simplified neutrino transport, 
it has been suggested that prompt convection increases the neutrinosphere 
luminosities (Burrows \& Fryxell 1993, Herant et al. 1994, Janka \& 
M\"{u}ller 1996) and pushes the shock out farther in radius (Janka 
\& M\"{u}ller 1996). Based on one-dimensional hydrodynamics, mixing-length 
convection, and detailed multigroup flux-limited diffusion ({\small MGFLD}) neutrino transport, 
all coupled self-consistently, it has been shown that prompt convection does 
not lead to a significant increase in the neutrino luminosities and has little 
effect on the shock dynamics (Bruenn \& Mezzacappa 1994, Bruenn et al. 1995). 
In this Letter, we remove the restriction to spherical symmetry in the
hydrodynamics and the need for mixing-length convection, and couple 
two-dimensional hydrodynamics to detailed one-dimensional {\small MGFLD} 
neutrino transport.

\section{Initial Models, Codes, and Methodology}

We began with the 25 ${\rm M}_{\odot}$ precollapse model S25s7b 
provided by Woosley (1995). The initial model was evolved through 
core collapse and bounce using {\small MGFLD} neutrino transport 
and Lagrangian hydrodynamics. The one-dimensional data at 8 ms 
after bounce (297 ms after the initiation of core collapse) were 
mapped onto our two-dimensional Eulerian grid. The inner and outer 
boundaries of our grid were chosen to be at radii 
of 25 km and 1000 km, respectively. 128 nonuniform radial spatial 
zones were used. This gave sufficient resolution at the inner 
boundary and at the shock. The nonuniform zoning was varied until 
no significant differences were seen between one-dimensional runs 
using 128 nonuniform and 512 uniform zones. 128 uniform angular zones 
spanning a range of 180 degrees and reflecting boundary conditions
were used for $\theta$. 

The inner boundary was chosen to be below the unstable region at
the onset of the simulation, which at 8 ms extended from 33 
to 58 km. At the base of the unstable region, $s=6.2$ and 
$Y_{\rm e}=0.23$, and at the top, $s=4.7$ and $Y_{\rm e}=0.16$,
where $s$ is the entropy per baryon in units of Boltzmann's 
constant and $Y_{\rm e}$ is the electron fraction. 
Spherically symmetric time-dependent boundary data for the 
two-dimensional hydrodynamics were supplied by our one-dimensional 
run. At each time step, the boundary data at our fixed inner 
Eulerian radius were extracted from the one-dimensional data 
by interpolation in $r$ and $t$. The outer boundary data were 
specified in the same way.

The two-dimensional hydrodynamics was evolved using an extended 
version, {\small EVH-1}, of the {\small PPM} hydrodynamics code 
{\small VH-1} developed by Blondin and colleagues at the Virginia 
Institute for Theoretical Astrophysics. Our extension allows for 
coupling to general equations of state. The matter in our simulations 
was in nuclear statistical equilibrium; to describe its thermodynamic 
state, we used the equation of state provided by Lattimer and Swesty 
(Lattimer \& Swesty 1991).

Because the finite differencing in our {\small PPM} scheme is
nearly noise free and because we cannot rely on machine roundoff
to seed convection in a time short compared with the hydrodynamics 
time scales in our runs, we seeded convection in the Ledoux unstable 
regions below and around the neutrinospheres by applying random 
velocity perturbations to the radial and angular velocities between 
$\pm$5\% of the local sound speed. 

In our two-dimensional simulations, gravity was assumed to be spherically 
symmetric. The gravitational field in the convectively unstable region 
was dominated by the enclosed mass at the region's base, which at the 
start of our simulations was 0.82 M$_{\odot}$. The enclosed mass at the 
top of the unstable region at this time was 1.1 M$_{\odot}$. Moreover, 
at $t=303$ ms in our simulation without neutrino transport (6 ms after 
the initiation of our run, at a time when prompt convection was fully 
developed), the density varied with $\theta$ about its average value 
(which was used in computing the spherically symmetric gravitational 
field) between -17\% and +24\% at 41 km and -9\% and +57\% at 50 km. 
Therefore, we do not believe this assumption was a serious shortcoming 
of our calculations. The time dependence of the mass enclosed by our 
inner boundary, given by our one-dimensional {\small MGFLD} runs, was 
taken into account. The solution of the Poisson equation for the 
gravitational potential will be incorporated in future investigations. 

The neutrino heating and cooling, and the change in the electron fraction, 
were computed using the following formulae: 
$d\epsilon /dt=c\sum_{i=1}^{2}\int E_{\nu}^{3}dE_{\nu}
[\psi^{0}_{i}/\lambda^{(a)}_{i}-j_{i}(1-\psi^{0}_{i})]/\rho (hc)^{3}$ 
and 
$dY_{\rm e}/dt=cm_{\rm B}\sum_{i=1}^{2}\alpha_{i}\int E_{\nu}^{2}dE_{\nu}
[\psi^{0}_{i}/\lambda^{(a)}_{i}-j_{i}(1-\psi^{0}_{i})]/\rho (hc)^{3}$, 
where $\epsilon$ is the internal energy per gram; 
$E_{\nu}$, $\psi^{0}_{i}$, $\lambda^{(a)}_{i}$, and $j_{i}$ are the 
electron neutrino or antineutrino energy, zeroth distribution function 
moment, absorption mean free path, and emissivity, respectively; 
$m_{\rm B}$ is the baryon mass; $i=1,\alpha_{1}=1$ corresponds to 
electron neutrinos, and $i=2,\alpha_{2}=-1$ corresponds to electron 
antineutrinos. The time-dependent $\psi^{0}_{i}$'s 
are obtained from tables in $r$ and $t$ constructed from our one-dimensional
{\small MGFLD} simulations. Comparisons were made to ensure that in one 
dimension the results from our code matched the results obtained with 
Bruenn's {\small MGFLD} code, modulo {\small EVH-1}'s better resolution 
of the shock.

%
In the 15 ${\rm M}_{\odot}$ case, we began with the precollapse model 
S15s7b provided by Woosley (1995). The one-dimensional data at 16 ms 
after bounce were mapped onto our two-dimensional Eulerian grid, and 
the inner and outer boundaries were chosen to be at radii of 20 km and 
1000 km, respectively. Initially, the unstable region extended from 29 
to 58 km, with $s$ and $Y_{\rm e}$ varying from 6.1 to 5.3 and 0.26 to
0.15, respectively. At this time, the enclosed mass at the bottom and 
top of the unstable region was 0.77 M$_{\odot}$ and 1.1 M$_{\odot}$, 
respectively. At $t=218$ ms in our simulation without neutrino transport 
(10 ms after the initiation of our run, at a time when prompt convection
was fully developed), the angular density variations about the average
density were between -15\% and +18\% at 41 km and -13\% and +65\% at 51 km.

\section{\bf Results}

For both our 15 and 25 M$_{\odot}$ models, two simulations were carried 
out for a duration of 100 ms after the initial postbounce time --- one 
with neutrino transport and one without. 

For our 25 M$_{\odot}$ model without transport, prompt convection develops and dissipates in 
$\sim$15 ms, by which time the initial gradients are smoothed out 
by convection. The entropy evolution is shown in Figure 1. Prolonged 
convection would require the maintenance of the electron fraction 
gradient by electron neutrino escape (neutrino transport) at the 
neutrinosphere. 
For the duration of the prompt convection episode, the fluid
velocity at the inner boundary remained positive, ensuring 
that none of this innermost convection was swept off our grid.

Prompt convection seeds unstable regions at successively larger 
radii and eventually convection reaches the shock and distorts 
it from sphericity. However, on the average, there is no significant 
difference between our one- and two-dimensional shock trajectories: 
convection does not significantly boost the stalled shock radius. 
[For details, see Calder et al. (1996b).] 

The outward motion of the shock in Figure 1 is not indicative of an 
explosion but results from the decreasing preshock accretion ram 
pressure and is seen in our one-dimensional run. 

With transport, both the convection growth rate and the asymptotic 
convection velocities are substantially smaller (see Section 4). The
asymptotic velocities become too small relative to the bulk inflow to 
result in any significant convective transport of 
entropy and leptons, and the evolution proceeds as it does in our 
one-dimensional run. The evolution is shown in Figure 2.

%
The results for the 15 M$_{\odot}$ model are essentially the 
same.
In the absence of neutrino transport, convection develops and
dissipates in $\sim$15 ms. Prompt convection seeds successive 
unstable regions between the prompt convection region and the 
shock. The shock is distorted from sphericity, but again, on 
the average, is not boosted significantly in radius relative 
to its position in our one-dimensional run without convection.

As in the 25 M$_{\odot}$ case, the prompt convection growth rate and 
asymptotic velocities in the presence of neutrino transport are too 
small to result in any significant convective transport of entropy and 
leptons. Complete details for both models will be given in Calder 
et al. (1996a). 

For our 15 M$_{\odot}$ model, the velocity at the inner boundary was 
negative during the course of the prompt convection episode; however, 
this did not have a significant effect on the development of convection 
in our ``hydrodynamics only'' run. Moreover, the key outcome, that 
neutrino transport inhibits the development of prompt convection, 
remained unchanged.

\section{Analytical Model}

Convection near or below the neutrinospheres can be profoundly 
influenced by the neutrino transport of energy and leptons between 
a convecting fluid element and the background. In effect, convection 
becomes ``leaky,'' and differences between a convecting fluid 
element's entropy and lepton fraction and the background's entropy
and lepton fraction, from which the buoyancy force driving convection 
arises, are reduced. 
To construct the simplest model of this, we will assume that the 
lepton fraction gradient is zero and that convection is driven 
by a negative entropy gradient that is constant in space and time. 
[Reversing the roles of the entropy and lepton fraction gradients 
would give analogous results. The more general case in which both 
gradients are nonzero has been considered 
by Bruenn and Dineva (1996). This complicates the analysis and can 
lead to additional modes of instability, such as semiconvection 
and neutron fingers, which are not relevant here.] We will also
assume that the effect of neutrino transport is to equilibrate 
the entropy of a fluid element with the background entropy in a 
characteristic time scale $\tau_{s}$. If the fluid element and 
the background are in pressure balance, and if we neglect viscosity, 
the fluid element's equations of motion are

\begin{equation}
\dot{v} = \frac{g}{\rho} \alpha_{s} \theta_{s}
\label{eq:zddot} 
\end{equation}

\begin{equation}
\dot{\theta}_{s} = - \frac{\theta_{s}}{\tau_{s}} - \frac{d\bar{s}}{dr} v
\label{eq:thetadot} 
\end{equation}

\noindent where $\theta_{s} = s - \bar{s}$, with $s$ and $\bar{s}$ being 
the fluid element's entropy and the background's entropy, respectively; $g$ is 
the local acceleration of gravity; $v$ is the radial velocity; and 
$\alpha_{s} \equiv - \left( \partial \ln \rho /\partial \ln s 
\right)_{P,Y_{\ell}} > 0$, where $Y_{\ell}$ is the common lepton 
fraction. Equation (\ref{eq:zddot}) equates the 
fluid element's acceleration to the buoyancy force arising from 
the difference between its entropy and the background's entropy; equation 
(\ref{eq:thetadot}) equates $\dot{\theta}_{s}$ to $\dot{s}$ minus 
$\dot{\bar{s}}$, where $\dot{s}$ results from the fluid element's 
equilibration with the background, and $\dot{\bar{s}}$ results from 
its motion through the gradient in $\bar{s}$. 

If we neglect neutrino effects ($\tau_{s} = \infty$), the solutions
to equations (\ref{eq:zddot}) and (\ref{eq:thetadot}) indicate that
(a) if $d\bar{s}/dr > 0$, the fluid element oscillates with the 
Brunt-V\"{a}is\"{a}l\"{a} frequency $\omega_{BV} \equiv \left[ 
(g\alpha_{s}d\bar{s}/dr)/\rho \right]^{1/2}$, and (b) if $d\bar{s}/dr 
< 0$, it convects, i.e., its velocity increases exponentially, and the 
convection growth time scale is given by $\tau = \tau_{BV} \equiv 
\left[ -(g\alpha_{s}d\bar{s}/dr)/\rho \right]^{1/2}$. 
When neutrino transport effects are included in the convectively 
unstable case ($d\bar{s}/dr < 0$), the fluid element convects,
but the convection growth time scale $\tau > \tau_{BV}$ is given by 
$1/\tau = \left[ 1/\tau_{BV}^{2} + 1/4\tau_{s}^{2} \right]^{1/2}- 1/2\tau_{s}$. 
In the limit $\tau_{s} \ll \tau_{BV}$, the growth time scale increases by 
$\tau_{BV}/\tau_{s}$, i.e., $\tau \simeq \tau_{BV}^{2}/\tau_{s}$. 

In addition to reducing convection's growth rate, neutrino transport 
also reduces its asymptotic velocities. In particular, in the limit 
$\tau_{s} \ll \tau_{BV}$, the solutions to equations (\ref{eq:zddot}) 
and (\ref{eq:thetadot}) show that a fluid element's velocity after 
moving a distance $\ell$ from rest is reduced by the factor $\tau_{s}/\tau_{BV}$. 

To apply this analysis to prompt convection, we note that $\tau_{BV}$ is 2 - 3 ms
in the region between $10^{11}$ and $10^{12}$ g cm$^{-3}$ in models S15s7b and S25s7b 
immediately after shock propagation. On the other hand, our $\dot{\epsilon}$ from 
neutrino heating and cooling gives values for $\tau_{s}$ that decrease from 
0.6 ms at $10^{11}$ g cm$^{-3}$ to 0.01 ms at $10^{12}$ g cm$^{-3}$. Our $\dot{Y}_{e}$ 
gives values for the lepton equilibration time scales that are about a factor of 5 smaller. 
These imply that neutrino transport should reduce the growth rate and asymptotic velocities 
of entropy-driven convection by factors of from $\sim 4$ near $10^{11}$ g cm$^{-3}$ to 
$\sim 250$ near $10^{12}$ g cm$^{-3}$. The lepton-driven convection growth rate and 
asymptotic velocities should be reduced by an additional factor of 5.

\section{\bf Caveat}

For model S15s7b, the Planck-averaged optical depth from the top
to the bottom of the unstable region varies from 1.2 to 24
for electron neutrinos and from 0.56 to 11.3 for electron
antineutrinos. The corresponding quantities for S25s7b vary
from 3.0 to 13.3 and from 1.4 to 5.7, respectively.

Because we are imposing a background neutrino distribution in 
our two-dimensional simulations, in optically thick regions
we are overestimating the rate for a fluid element to equilibrate 
with the background, and therefore, overestimating the effect 
transport has on inhibiting the development of prompt convection. 
The equilibration would be affected by the advected trapped 
neutrinos and by the finite time for neutrino transport between 
the element and its surroundings, both of which are neglected in 
our analysis.

Equilibration experiments (Bruenn et al. 1995, Bruenn \& Dineva 1996),
which include these effects, show that a fluid element of one pressure 
scale height in radius will equilibrate in entropy with a time scale of 
0.45 ms at $3\times 10^{11}$ g cm$^{-3}$ and 1 ms at $1\times 10^{12}$ 
g cm$^{-3}$. Smaller modes will equilibrate faster. The equilibration 
time scale for $Y_{\rm e}$ is 0.1 - 0.4 ms at $3\times 10^{11}$ g cm$^{-3}$ and 
0.2 - 1 ms at $1\times 10^{12}$ g cm$^{-3}$. These time scales are small 
compared with the Brunt-V\"{a}is\"{a}l\"{a} time scales --- implying that
our conclusions are valid --- but not as small as our heating and cooling 
rates predict.

\section{\bf Summary}

Near and below the neutrinosphere, neutrino transport equilibrates 
a convecting fluid element with its surroundings in both entropy and 
electron fraction in a fraction of a millisecond. As a result, prompt 
convection growth rates and asymptotic velocities are reduced by factors 
of 4--250. Prompt convection velocities become too small relative to 
the bulk inflow to result in any significant convective 
transport of entropy and leptons to the neutrinospheres; therefore, prompt 
convection will have no effect on boosting the neutrinosphere luminosities 
nor on boosting the neutrino reheating of the stalled supernova shock wave.

\section{Acknowledgements}

AM, ACC, MWG, and MRS were supported at the Oak Ridge National 
Laboratory, which is managed by Lockheed Martin Energy Research 
Corporation under DOE contract DE-AC05-96OR22464. AM, MWG, and MRS were 
supported at the University of Tennessee under DOE contract 
DE-FG05-93ER40770. ACC and SU were supported at Vanderbilt 
University under DOE contract DE-FG302-96ER40975. SWB was 
supported at Florida Atlantic University under NSF grant 
AST--941574, and JMB was supported at North Carolina State 
University under NASA grant NAG5-2844. 
The simulations presented in this Letter were carried out on
the Cray C90 at the National Energy Research Supercomputer
Center, the Cray Y/MP at the North Carolina Supercomputer 
Center, and the Cray Y/MP and Silicon Graphics Power
Challenge at the Florida Supercomputer Center. We would 
like to thank Thomas Janka and Friedel Thielemann for 
stimulating discussions.

\section{\bf References}

\begin{enumerate}
\item Bethe, H. \& Wilson, J. R. 1985, ApJ, 295, 14
\item Bruenn, S. W. \& Mezzacappa, A. 1994, ApJ, 433, L45
\item Bruenn, S. W., Mezzacappa, A., \& Dineva, T. 1995, Phys. Rep., 256, 69
\item Bruenn, S. W. \& Dineva, T. 1996, ApJ, 458, L71
\item Burrows, A., Hayes, J., \& Fryxell, B. A. 1995, ApJ, 450, 830
\item Burrows, A. \& Fryxell, B. A. 1993, ApJ, 418, L33
\item Calder, A. C., Mezzacappa, A., Bruenn, S. W., Blondin, J. M., 
      Guidry, M. W., Strayer, M. R., \& Umar, A. S. 1996a, ApJ, in preparation
\item Calder, A. C., Mezzacappa, A., Bruenn, S. W., Blondin, J. M., 
      Guidry, M. W., Strayer, M. R., \& Umar, A. S. 1996b, ApJ, in preparation
\item Herant, M. E., Benz, W., Hix, W. R., Fryer, C., \& Colgate, S. A. 1994, 
      ApJ, 435, 339
\item Herant, M. E., Benz, W., \& Colgate, S. A. 1992, ApJ, 395, 642
\item Janka, H.-Th., \& M\"{u}ller, E. 1996, A\& A 306, 167 
\item Lattimer, J. M. \& Swesty , F. D. 1991, Nucl. Phys. A, 535, 331
\item Mezzacappa, A., Calder, A. C., Bruenn, S. W., Blondin, J. M., 
      Guidry, M. W., Strayer, M. R., \& Umar, A. S. 1996, ApJ, in preparation
\item Miller, D. S., Wilson, J. R., \& Mayle, R. W. 1993, ApJ, 415, 278
\item Wilson, J. R. \& Mayle, R. W. 1993, Phys. Rep., 227, 97 
\item Wilson, J. R. 1985, in Numerical Astrophysics, eds. J. M. Centrella et al.
      (Boston: Jones \& Bartlett), 422 
\item Woosley, S. E. 1995, private communication
\end{enumerate}

\newpage

\figcaption{The 25 M$_{\odot}$ model: without neutrino transport, the evolution
in entropy shows the development of prompt convection and the convection it seeds 
between the prompt convection region and the shock.}

\figcaption{The 25 M$_{\odot}$ model: with neutrino transport, the evolution in 
entropy shows no significant convection.}

\end{document}